# Electronic Decoupling of Polyacenes from the Underlying Metal Substrate by $sp^3$ Carbon Atoms

*Mohammed S. G. Mohammed,*[1,2,‡] *Luciano Colazzo,*[1,2,‡,†] *Roberto Robles,*[2,*] *Ruth Dorel,*[3,⊥] *Antonio M. Echavarren,*[3,4] *Nicolás Lorente,*[1,2] *Dimas G. de Oteyza*[1,2,5,*].

[1] Donostia International Physics Center (DIPC), 20018 San Sebastián, Spain

[2] Centro de Física de Materiales (CSIC-UPV/EHU) – MPC, 20018 San Sebastián, Spain

[3] Institute of Chemical Research of Catalonia (ICIQ), Barcelona Institute of Science and Technology, 43007 Tarragona, Spain

[4] Departament de Química Orgànica i Analítica, Universitat Rovira i Virgili, 43007 Tarragona, Spain

[5] Ikerbasque, Basque Foundation for Science, 48013 Bilbao, Spain

KEYWORDS. on-surface synthesis; heptacene, pentacene; scanning tunneling microscopy; scanning tunneling spectroscopy; density functional theory

ABSTRACT. We report on the effect of $sp^3$ hybridized carbon atoms in acene derivatives adsorbed on metal surfaces, namely decoupling the molecules from the supporting substrates. In particular, we have used a Ag(100) substrate and hydrogenated heptacene molecules, in which the longest conjugated segment determining its frontier molecular orbitals amounts to five consecutive rings. The non-planarity that the $sp^3$ atoms impose on the carbon backbone results in electronically decoupled molecules, as demonstrated by the presence of charging resonances in dI/dV tunneling spectra and the associated double tunneling barriers, or in the Kondo peak that is due to a net spin S=1/2 of the molecule as its LUMO becomes singly charged. The spatially dependent appearance of the charging resonances as peaks or dips in the differential conductance spectra is further understood in terms of the tunneling barrier variation upon molecular charging, as well as of the different orbitals involved in the tunneling process.

TOC

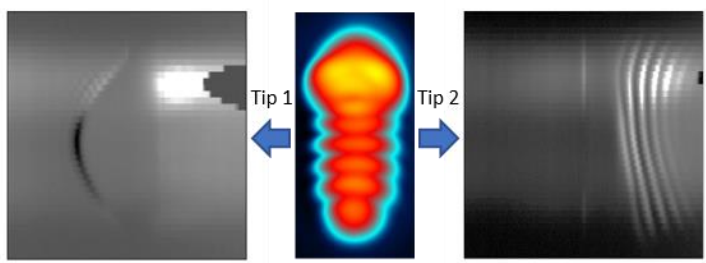

Polyaromatic hydrocarbon (PAH) molecules attract great interest due to the enormously tunable properties they display as a function of their chemical structure. For example, by rearranging their carbon backbones, PAHs can change from wide bandgap organic semiconductors to high-spin organic ferromagnets.[1] In the particular case of polyacenes, a subclass of PAHs that consist of linearly fused benzene rings, a notable size-dependence is observed, changing from a closed-shell to a dominant open-shell character as the number of rings increases beyond six.[2,3] The latter confers a high reactivity upon such acenes. To circumvent the associated difficulties in their synthesis, the surface-supported synthesis under ultra-high-vacuum (UHV) conditions[4] has become an increasingly popular approach.[5–15]

The vacuum environment and the atomically flat substrates further provide excellent conditions for molecular characterization, which can be performed at the single molecule level by scanning probe microscopy. However, when adsorbed on metallic surfaces, hybridization of the molecules with the substrate significantly modifies their properties and leads, at best, to a substantial broadening of the orbitals. Thus, a decoupling of the molecules from the underlying metal substrate becomes not only beneficial but often strictly necessary to characterize the intrinsic molecular properties.

To this aim, a variety of methods have been applied. Examples include the intercalation of new species between metal and adsorbate (extensively used e.g. in graphene-based research),[16–18] or the use of metal-supported insulating buffer layers as substrates.[19–21] The chemical addition of bulky side groups to the molecular structure of interest, which can act as molecule-substrate spacers and thereby reduce their coupling strength, has also been proved successful.[22,23] Herein we report the decoupling of polyacenes from an underlying Ag(100) surface by exploiting the non-planarity of the organic backbone imposed by *sp*$^3$-type functional groups.

Hydrogenated polyacene derivatives are stable molecules that can be easily deposited onto clean surfaces by sublimation under ultra-high-vacuum conditions and have been used as starting reactants for their transformation into acenes by controlled dehydrogenation.[10,11,13] As can be discerned in Fig. 1a and Fig. 1b, the hydrogenated rings (imaged by STM with CO-functionalized probes in the repulsive regime as larger rings (Fig. 1c,d)) feature $sp^3$ hybridized carbon atoms that break the conjugation along the molecule. It has been demonstrated that the energy gap between the frontier electronic states (highest occupied molecular orbital/HOMO and lowest unoccupied molecular orbital/LUMO) of such molecules is ultimately determined by the longest conjugated acene segment, and is remarkably similar to the bandgap of acenes of the same length.[10,11,13] That is, the di-hydrogenated heptacene derivative displayed in Fig. 1b,d, which has been obtained from the controlled tip-induced dehydrogenation of 5,9,14,18-tetrahydroheptacene precursors (Fig. 1a,c),[10] features five rings as the longest conjugated segment and is characterized by a bandgap similar to that of pentacene.[10] Proof of the similar electronic properties of 5,18-dihydroheptacene (hereafter referred as dihydroheptacene) and pentacene is provided in Figs. 1e-h, which display calculated low-energy states for the molecules adsorbed on Ag(100). In the case of pentacene they strongly resemble the LUMO (Fig. 1e) and HOMO (Fig. 1g) of free-standing pentacene,[21] while the analogous calculations for dihydroheptacene (Fig. 1f,h) reveal a striking similarity. As such, dihydroheptacene can also be considered as a "functionalized pentacene" (2,3-alkylpentacene).

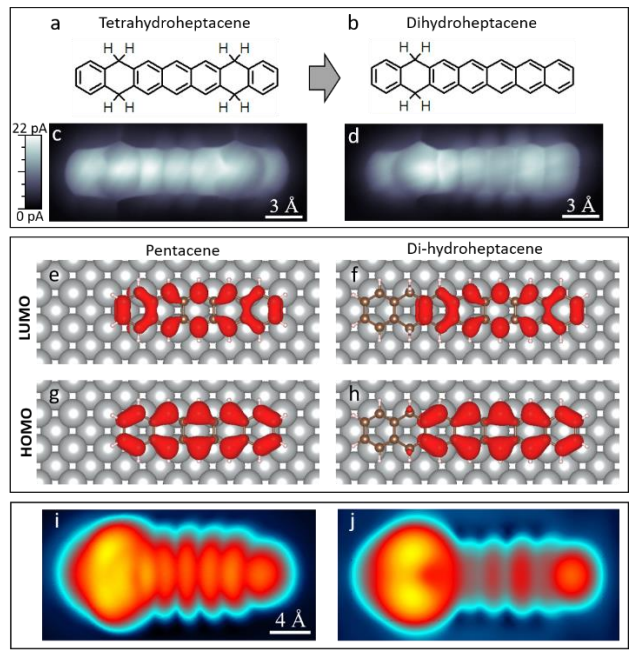

**Figure 1**. Chemcal structure of 5,9,14,18-tetrahydroheptacene (a) and 5,18-dihydroheptacene (b). Constant height STM images with CO-functionalized probes in the repulsive interaction regime of 5,9,14,18-tetrahydroheptacene (c) and 5,18-dihydroheptacene (d) (U = 2 mV). Calculated charge densities associated with the frontier molecular orbitals of pentacene (LUMO (e) / HOMO (g)) and 5,18-dihydroheptacene (LUMO (f) / HOMO (h)). Constant current STM image of 5,18-dihydroheptacene recorded at U=20 mV / I=100 pA (i) and a simulated STM image of the 5,18-dihydroheptacene LUMO level (j). Note that the differences between experiment (i) and theory (j) arise from the *s*-wave probe used in the simulations *vs*. the *p*-wave character of the CO-functionalized probe used in the measurements.[21]

Previous studies of pentacene on Ag(100) have shown the energy level alignment to be such that the LUMO is aligned with the Fermi energy and consequently partially occupied with an estimated charge of 0.7 electrons.[24] Any charge transfer-related magnetism is hindered by the substrate: The LUMO is substantially broadened by hybridization with the metal, and the high level of screening reduces electron correlations.[25,26] In contrast, when the molecule is electronically decoupled from the underlying Ag(100) by a MgO bilayer, the integer charge transfer of one electron to pentacene confers the system a net spin S=1/2 and two spin-split resonances of the former LUMO above and below the Fermi level (the singly occupied (SOMO) and singly unoccupied (SUMO) molecular orbitals, respectively).[24]

Focusing now on dihydroheptacene on Ag(100), the contrast in scanning tunneling microscopy (STM) measurements at low bias values close to the Fermi level (Fig. 1i) resembles the simulated image of the molecular LUMO (Fig. 1j). This underlines again the similarities of dihydroheptacene with pentacene in terms of molecular bandgap and energy level alignment, with the LUMO energy around $E_F$. However, it is important to note that there is a great variability in the scanning tunneling spectroscopy (STS) measurements of dihydroheptacene. Figure 2 shows some examples thereof (further examples are displayed in Fig. S1). Panels b, c and d display color-coded stacks of dI/dV point-spectra taken along a molecule as marked by the yellow arrow in Fig. 2a. Sample spectra corresponding to the dashed lines (and to the position marked in Fig. 2a) are shown below. The datasets of Fig. 2b, 2c and 2d correspond to different molecules, although all of them share the same chemical structure displayed in Fig. 1b. Beyond their markedly different appearance, all datasets present sequences of sharp peaks that shift in position along the molecule. The narrow width of the peaks, their observation beyond the limits of the molecule (i.e. near the molecule but on the bare substrate), the electric field dependence of their energy alignment (Fig. S2) and their ring-like appearance in conductance maps (Fig. 3c), all together provide unambiguous proof of their nature, namely so-called charging resonances (and associated vibronic resonances[27]) related to a double tunneling barrier.[27,28] The presence of a double tunneling barrier implies an effective decoupling of the molecule from the underlying substrate, which is neither expected from molecules in direct contact with transition metal surfaces nor observed for the closely related pentacene on Ag(100). Here, the key to rationalizing this surprising finding is the presence of the hydrogenated $sp^3$ C atoms of dihydroheptacene, which impose a non-planarity to the molecule (Fig. 2e) that sufficiently reduces its electronic coupling from the substrate to generate the double barrier in the tunneling process.

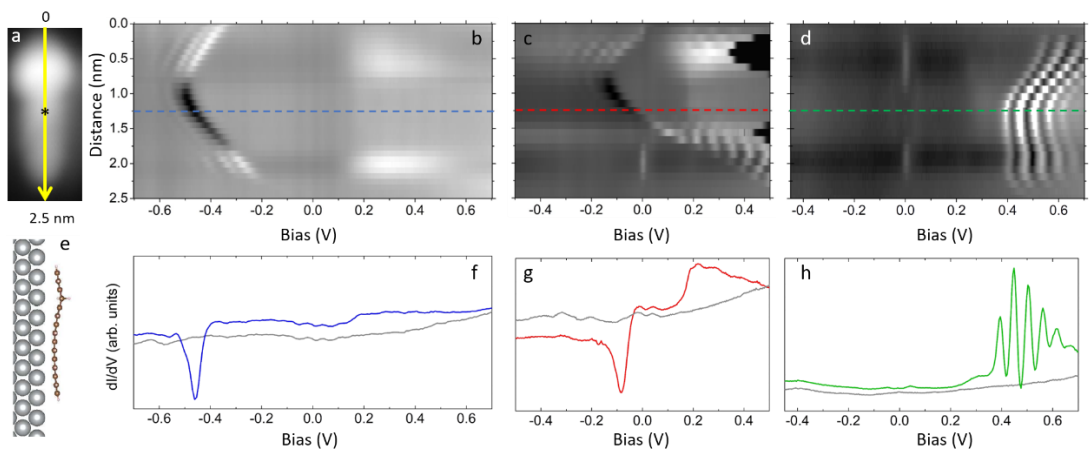

**Figure 2.** (a) Constant current image of 5,18-dihydroheptacene, marking with the yellow line the typical trajectory along which the point spectra are acquired that make up the following panels. (b-d) Stacked point spectra with color-coded dI/dV signal (at each point the feedback loop is adjusted to U=200 mV / I=200 pA (b), U=50 mV / I=250 pA (c) and U=50 mV / I=100 pA (d) ). The spatial coordinate in the vertical axis is plotted to approximately coincide with the yellow line in panel (a). Panels (b), (c) and (d) correspond to measurements on three different molecules sharing the same structure. (e) Relaxed 5,18-dihydroheptacene structure on Ag(100) (as obtained from DFT) evidencing a clear non-planarity. (f-h) Selected dI/dV point spectra extracted from panels (b-d) at the position of the dashed lines and of the asterisk in panel (a). The corresponding reference spectra measured on the Ag(001) surface next to the molecules are displayed by grey lines.

In a conventionally pictured double barrier, at zero sample bias there is no electric field across the tunneling junction (Fig. S3a,b). When a positive sample bias is applied, an electric field is established and the associated potential drops partially across the first (substrate-adsorbate) barrier (Fig. S3b). As a consequence, the molecular orbitals shift up in energy with respect to the substrate´s Fermi level. At a certain bias threshold, an occupied orbital can eventually cross the Fermi level and cause the transfer of an electron to the substrate. If the tip-sample distance (second tunneling barrier) is reduced, the voltage drop across the tip-molecule barrier is enhanced, requiring lower bias values to reach the charge transfer threshold (Fig. S3c). The opposite scenario would apply at negative sample bias: The molecular orbitals would be shifted to lower energies until an electron is transferred from the substrate to an initially unoccupied molecular orbital at a threshold bias that decreases (in absolute numbers) as the tip-sample distance is reduced.

An interesting point is that in our measurements the charging resonances shift in the same direction (to lower sample bias values) when reducing the tip-sample distance (Fig. S2) and display the same concave curvature in the stacked spectra regardless of whether they appear at positive (Fig. 2d) or negative (Fig. 2b) bias. In fact, the curved charging resonance trajectory along the molecule often crosses the Fermi level (Fig. 2c). This is inconsistent with the common assumption of a zero electric field at zero bias and instead implies that, at least in cases like Fig. 2b and Fig. 2c, the zero electric field condition occurs at negative sample bias, in particular more negative than the "charging peak parabola" onset (Fig. S3d). This is understood in a straightforward manner assuming different work functions for tip and sample, which establish a contact potential difference and the associated electric field when brought into tunneling distance (Fig. S3d). Indeed, the zero electric field bias corresponds to that typically measured in Kelvin Probe Force Microscopy, which in most of the cases differs from zero and varies with both tip and sample.[29]

We can thus now conclude that the charging resonance involves in all cases (whether at positive or negative bias values) the same molecular orbital, namely the LUMO. It is close in energy to $E_F$ and can thus be easily brought across the substrate´s Fermi level by electric fields at the tunneling junction. At bias values above or below that of the charging resonance the LUMO is empty or occupied, respectively. For that reason, whenever the zero bias point coincides with a molecule in its charged state (occupied LUMO), a zero bias resonance appears in the spectra (Fig. 2c-d), but not if the molecule at zero bias is in its neutral state (Fig. 2b-c). The zero bias resonance is a manifestation of the Kondo effect associated with the spin 1/2 of the molecule as a single electron occupies its gas-phase LUMO.[30,31] Its temperature-dependent width is in agreement with a Kondo temperature of 73±2 K (Fig. S4) and its presence is further proof of the decoupling effect that the "pentacene functionalization" at C-2 and C-3 with $sp^3$ carbon atoms brings about. If that were not the case, a broader LUMO level and reduced electron correlations would cause an equal population of spin up and spin down LUMO orbitals, resulting in no net spin and no Kondo peak, as in pristine pentacene.[24]

A question that arises at this point, however, is where the large variability displayed by the charging resonances comes from. Being an electric field-driven effect, it depends largely on the tip, on the sample, as well as on the contact potential difference between them. Different tips (in terms of shape, but also tip apex functionalization)[29] will notably impact the electric field lines within the tunneling junction and thus the charging resonance appearance. By way of example, Fig. 3a shows the conductance spectrum obtained with a metallic tip on the "functionalized pentacene" molecule marked with a blue point in the inset. The charging peak appears at negative bias. Consequently, at zero bias the molecule is in its neutral state and no Kondo peak is observed. As the tip is functionalized by picking up the molecule marked with the white oval, the tip´s work function and the overall electric field at the tunneling junction change dramatically. As a result, a conductance spectrum taken on the same molecule as in panel (a) now shows the charging resonances at positive bias (Fig 3b). The molecule is therefore charged at zero bias and displays a Kondo peak.

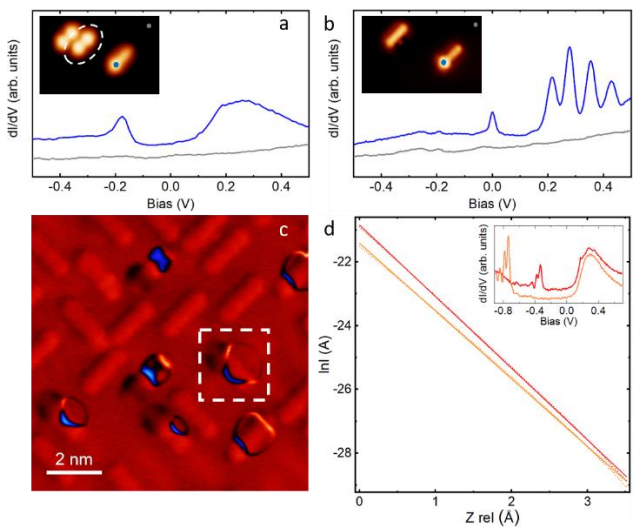

**Figure 3**. (a) dI/dV point spectrum on a 5,18-dihydroheptacene molecule (blue line), along with the corresponding reference spectrum on the substrate (grey line). The respective positions are marked with correspondingly colored dots in the inset. (b) Similar dI/dV point spectra taken at the same positions (marked again with the colored dots in the inset) after functionalizing the tip with the 5,9,14,18-tetrahydroheptacene molecule marked with the white oval in the inset of panel (a). (c) Constant current dI/dV map (U=-80 mV, I=200 pA, Lock-in $U_{osc}$=1 mV) of a sample containing multiple molecular species including several 5,18-dihydroheptacene molecules, all of which display differently shaped charging resonances. (d) STS I vs z spectra on two different 5,18-dihydroheptacene molecules. The inset displays the associated dI/dV spectra on those same molecules.

However, not only the tip matters for the charging resonance appearance, but also inhomogeneities in the sample. The tunneling barrier and the electric field within the tunneling junction is also affected by the polarizability of a molecule´s surroundings, by neighboring molecules, as well as by surface and subsurface defects.[31–33] That is, even with the same tip, molecules sharing the same structure may display different charging peaks.[31] Fig. 3c displays a conductance map measured at -80 mV of a variety of hydrogenated heptacene derivatives on Ag(100).[10] The charging rings involving the "functionalized pentacene" molecules all show different shapes, sizes, and are even centered on different molecular positions, underlining the aforementioned variability. Fig. 3d shows STS data acquired with the same tip on equivalent positions of different "functionalized pentacene" molecules. The inset shows the dI/dV spectra, revealing charging resonances at very different bias values, around 0.4 V apart. Associated I vs. z curves taken at the same positions reveal the expected linear behavior when the current is plotted logarithmically, and notably different slopes (Fig. 3d). The slopes correlate with the effective barrier at the tunneling junction, which is typically associated with an average of tip and sample workfunction.[34] The tip being the same and other parameters like the geometry or local density of states being virtually unchanged between single molecules on the same substrate, the change in the effective barrier must relate to local substrate inhomogeneities. The effective barrier height associated with the red curve is 4.81 eV, while that of the orange curve is 0.51 eV lower. As discussed above, the zero electric field occurs at bias values below that of the charging peak, implying a lower work function of the substrate than of the tip. For the red curve, the higher barrier implies a larger substrate work function, thus closer to that of the tip. In turn, the electric field at zero bias is lower and causes a smaller upward shift in energy of the LUMO. As a result, the charging peak appears at lower bias values.

When the charging resonances occur at positive bias, they appear everywhere as a sharp maximum (Fig. 2d). Interestingly, the charging resonances at negative bias instead show a location-dependent appearance, namely as a peak near the molecular ends and as a dip around the "pentacene" segment's center or on the substrate (Fig. 2b,c). The same effect can be observed in conductance maps of the charging rings. When they appear at positive bias, the charging rings (and their concentric vibronic satellites) are imaged everywhere as a conductance increase (Fig. 4a,b). Instead, charging rings at negative bias are imaged with a reduced conductance signal on the substrate, an even stronger reduction around the "pentacene" segment's center, and with an increased conductance signal near its ends (Fig. 4c,d). We rationalize these findings as follows.

As an adsorbate becomes charged with an electron (hole) from the surface, that charge redistribution causes an interfacial dipole that increases (decreases) the effective tunneling barrier. In our experiments, at bias values below that of the charging resonance we have molecular species charged with an extra electron. This charge transfer causes an interface dipole that increases the tunneling barrier and thus reduces the associated tunneling current. As the molecule becomes neutral at higher bias values, the barrier decreases and the tunneling current increases. While at positive bias values this increase in current results in a sharp peak in the dI/dV signal (Fig. 4e), a similar increase of the tunneling intensity at negative bias appears as a dip (Fig. 4f). Taking these considerations into account we would expect the charging resonances to appear everywhere as dips at negative bias and as peaks at positive bias.

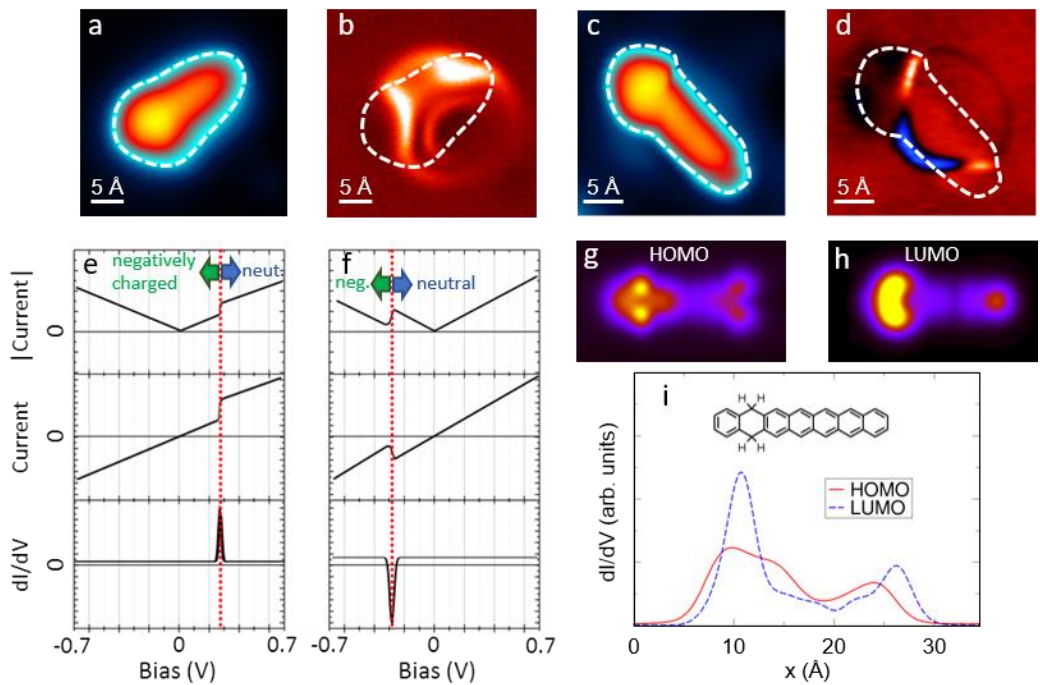

**Figure 4**. Topography (a,c) and simultaneously acquired dI/dV signal (b,d) of a 5,18-dihydroheptacene molecule displaying the charging resonances at positive (a,b; U=615 mV, I=80 pA) and negative bias (c,d; U=-80 mV, I=200 pA, Lock-in $U_{osc}$=1 mV). The latter corresponds to the molecule marked with a dashed square in Fig. 3c. Note that the ring's asymmetry, shape and center vary from molecule to molecule. Schematic graphs of the expected current modulus, current and dI/dV signal evolution associated with the local work function modification as a molecule changes its charge state at positive (e) and negative bias (f). Simulated dI/dV maps at 10 Å above the molecular plane at the HOMO (g) and LUMO (h) energies. (i) Comparative cross section of the calculated dI/dV signal along the long molecular axis at 10 Å above the molecular plane (the molecular structure inset is drawn to scale for guidance).

However, there is an additional effect to consider. In a singly charged molecule, the lowest energy states are the singly occupied molecular orbital (SOMO) and the singly unoccupied molecular orbital (SUMO), both of them related to the gas-phase LUMO. Instead, for a neutral molecule the frontier orbitals are the HOMO and the LUMO. That is, the unoccupied frontier states for charged and neutral molecules share the same wavefunction symmetry and spatial distribution of the gas-phase LUMO. Thus, for positive bias values on either side of the charging resonance, the orbitals involved in the tunneling (whether resonantly or only through the electronic state's tails) are always LUMO-related.

Focusing on the occupied states, the frontier states for charged and neutral molecules relate to the gas-phase LUMO and HOMO, respectively. Thus, for negative bias values on either side of the charging resonance, two different orbitals with very disparate wavefunction symmetry and spatial distribution are involved in the tunneling process. Their decay perpendicular to the surface plane (Fig. 4g,h) and consequently the tunneling transmission function between tip and sample is locally very different. As a result, at the same tip height it is the HOMO that provides a larger signal in the central region of the "functionalized pentacene", while it is the LUMO on the outer sides of the molecule (Fig. 4i). This has a strong impact on the measured conductance as the charging resonance is crossed while ramping up the bias and the orbital dominating the tunneling process changes from the gas-phase LUMO to the gas-phase HOMO: the conductance signal increases towards the molecular ends and decreases towards the center of the "pentacene" segment.

This effect is convoluted with that described earlier of the different tunneling barrier at either side of the charging resonance. At negative bias, the latter imprints a dip-like appearance to the charging resonances on the substrate nearby the molecules (Fig. 4d). As the tip moves onto the molecules and the molecular orbitals become directly involved in the tunneling process, the modified tunneling transmission promotes the dip on the central region of the molecule (Fig. 4d). In contrast, on the outer regions of the molecule it reduces the current as the charging threshold is crossed, compensating and even reverting the effect of the lower effective tunneling barrier on neutral molecules, thus finally appearing as a dI/dV peak in those regions.

In conclusion, we have characterized the critical effect of $sp^3$ carbon atoms along the backbone of acene derivatives on their electronic decoupling from the supporting substrates. Partially hydrogenated heptacene molecules, in which the longest conjugated segment determining its electronic properties corresponds to a "pentacene", have been used as test molecules. The non-planarity imposed by the $sp^3$ carbon atoms drives an electronic decoupling from the underlying Ag(100) substrate that has been shown to be sufficient to establish a double tunneling barrier and stabilize a singly charged molecule holding a net spin S=1/2. Interestingly, the charging resonances display a spatially dependent appearance as peaks or dips, which has been associated with the tunneling barrier variation upon molecular charging, as well as with the different orbitals contributing to the electron tunneling in their charged or neutral state.

ASSOCIATED CONTENT

**Supporting Information**. Additional stacks of dI/dV point spectra across different molecules or the same molecule with different tips, acquired without readjustment of the tip height between point and point (Fig. S1), sample dI/dV spectra evidencing the charging resonance shift for varying tip heights (Fig. S2), schematic diagrams of the energy level alignment and the tunneling barriers between tip and sample for different scenarios (Fig. S3), higher resolution spectra of the Kondo resonance and analysis of its temperature-dependent width (Fig. S4). (PDF)

AUTHOR INFORMATION

**Corresponding Author**

* roberto.robles@ehu.eus; * d_g_oteyza@ehu.eus

**Present Addresses**

† Center for Quantum Nanoscience, Institute for Basic Science (IBS), Seoul 03760, Republic of Korea and Department of Physics, Ewha Womans University, Seoul 03760, Republic of Korea.

ᴸ Stratingh Institute for Chemistry, Zernike Institute for Advanced Materials, University of Groningen Nijenborgh 4, 9747AG Groningen, The Netherlands

**Author Contributions**

The manuscript was written through contributions of all authors. All authors have given approval to the final version of the manuscript. ‡These authors contributed equally.

ACKNOWLEDGMENT

This project has received funding from the European Union's Horizon 2020 research and innovation programme under Grant Agreement Nos. 635919 (ERC-StG), 837225 (ERC-PoC) and 766864 (FET-Open), from the Spanish MINECO (Grant No. MAT2016-78293-C6), from AGAUR (2017 SGR 1257), and from the CERCA,Program / Generalitat de Catalunya.

# Supplementary Information

# Electronic Decoupling of Polyacenes from the Underlying Metal Substrate by $sp^3$ Carbon Atoms

*Mohammed S. G. Mohammed,*[1,2,‡] *Luciano Colazzo,*[1,2,‡,†] *Roberto Robles,*[1,2] *Ruth Dorel,*[3,⌐]
*Antonio M. Echavarren,*[3,4] *Nicolás Lorente,*[1,2] *Dimas G. de Oteyza*[1,2,5,*].

[1] Donostia International Physics Center (DIPC), 20018 San Sebastián, Spain

[2] Centro de Física de Materiales (CSIC-UPV/EHU) – MPC, 20018 San Sebastián, Spain

[3] Institute of Chemical Research of Catalonia (ICIQ), Barcelona Institute of Science and Technology, 43007 Tarragona, Spain

[4] Departament de Química Orgànica i Analítica, Universitat Rovira i Virgili, 43007 Tarragona, Spain

[5] Ikerbasque, Basque Foundation for Science, 48013 Bilbao, Spain

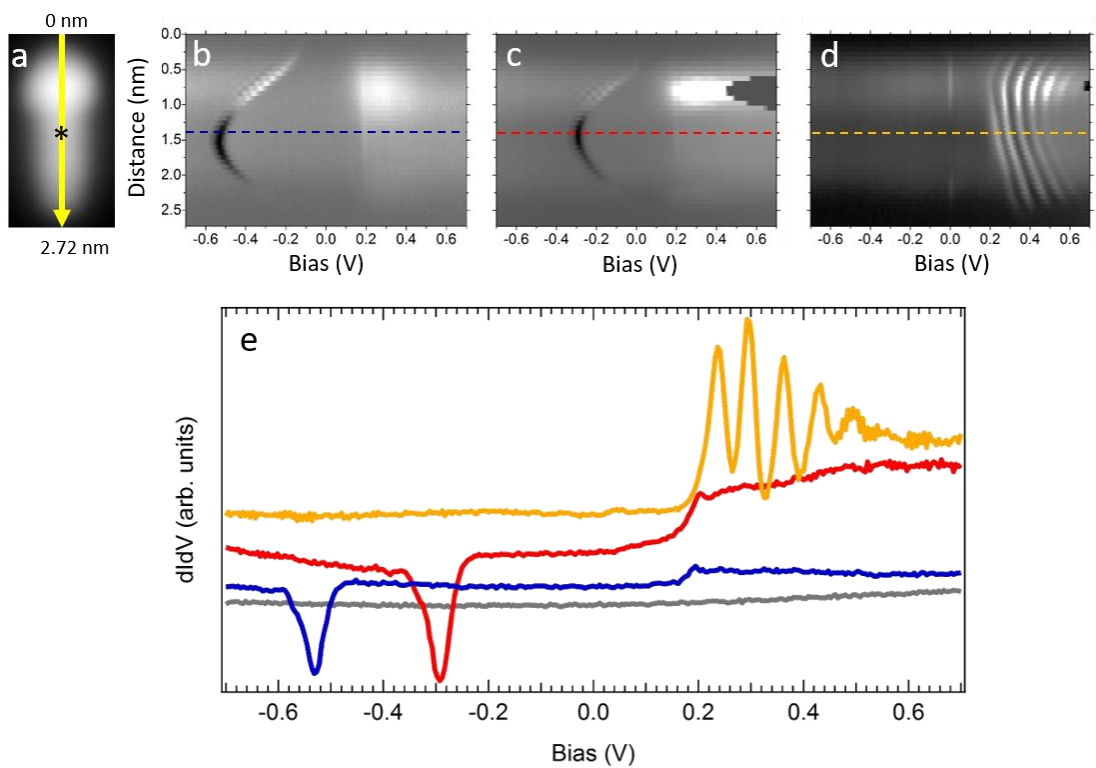

**Figure S1.** (a) Constant current image of dihydroheptacene, marking with the yellow line the typical trajectory along which the point spectra are acquired that make up the following panels. (b-d) Stacked point spectra with color-coded dI/dV signal obtained at constant height. The spatial coordinate in the vertical axis is plotted to approximately coincide with the yellow line in panel (a). Panels (b) and (c) correspond to measurements on two different molecules sharing the same structure, while panels (c) and (d) correspond to data acquired on the same molecule with two different tips. (e) Selected dI/dV point spectra extracted from panels (b-d) at the position of the correspondingly colored dashed lines and of the asterisk in panel (a), along with a typical reference spectrum measured on the Ag(001) surface next to the molecules (grey line).

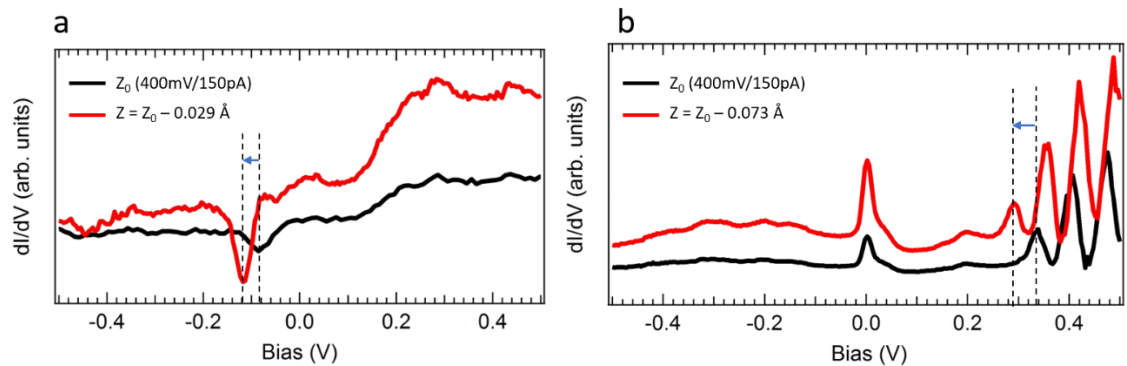

**Figure S2.** Selected point spectra displaying the charging resonances at negative (a) and positive (b) bias values, in each case measured at two different tip-sample distances. Regardless of whether the charging resonance appears at negative or positive bias, it shifts to lower energies as the tip-sample distance is reduced.

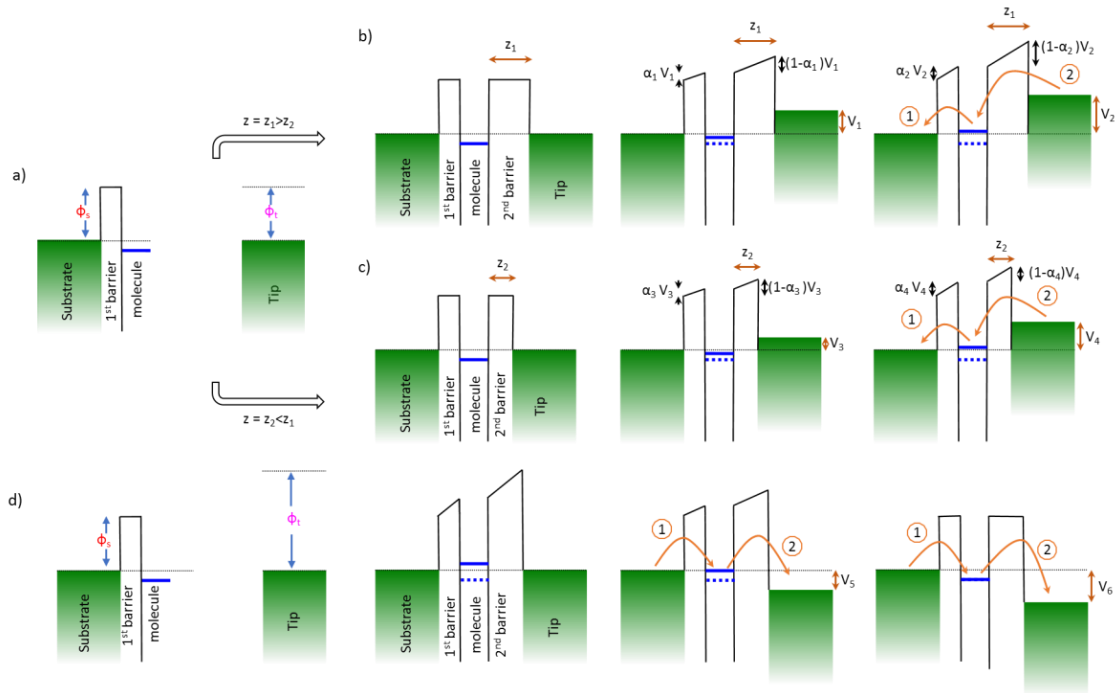

**Fig. S3.** Schematic diagrams of the energy level alignment and the tunneling barriers between tip and sample for different scenarios: (a) Separate tip and sample (consisting of substrate and molecule separated by a decoupling barrier), both with the same workfunction. (b) Tip and sample of similar work function brought into tunneling distance at zero bias, at a bias $V_1$ that brings an originally occupied molecular state to an energy close to $E_F$, and a at a bias $V_2$ that lifts the energy of the originally occupied molecular state above $E_F$. This causes an initial electron transfer from the molecule to the substrate, which subsequently opens up a tunneling channel from the tip to the now-empty molecular state. (c) Similar scenario as described in (b) but with a smaller tip-sample distance $z_2$, which enhances the voltage drop at the substrate-molecule barrier and thus lowers the voltage $V_4$ required to trigger the charge transfer from molecule to substrate. (d) Energetics when bringing into tunneling distance a sample to a tip with a larger workfunction. An electric field is readily created at zero bias that may bring an originally occupied molecular orbital above $E_F$ and trigger electron transfer from the molecule to the substrate. The threshold bias for the charge transfer is indeed at a negative bias $V_5$, and the actual zero electric field condition occurs at an even more negative bias value $V_6$.

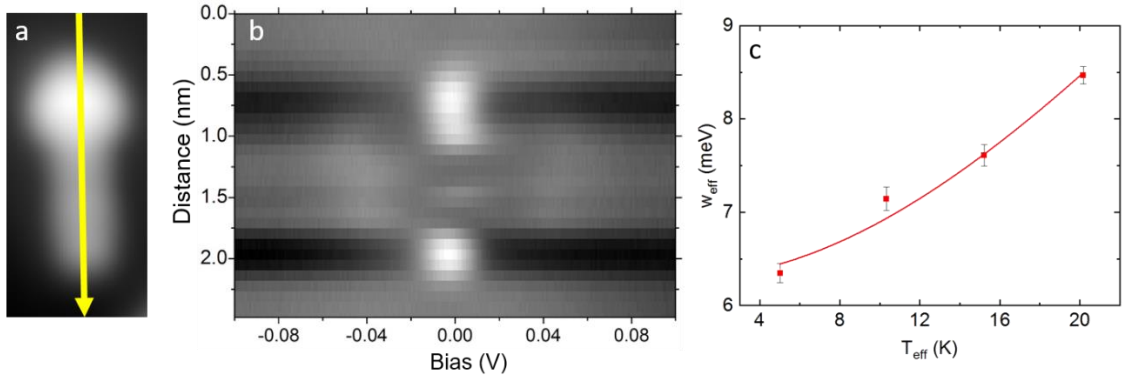

**Fig. S4.** Constant current image of dihydroheptacene (a), marking with the yellow line the trajectory along which the point spectra are acquired that make up panel (b) when stacked into a color-coded map. (c) Effective width of the zero-bias resonance as a function of the effective temperature (see details below) and a fit to the Fermi-Liquid model (solid line) corresponding to a Kondo temperature of 73±2 K.

The Kondo temperature analysis is performed following the procedure described in [1]. In a first step, we correct for the broadening due to the finite tip temperature as proposed by Zhang et al. [2]: $w_{eff} = \sqrt{\Delta^2 - \Delta_{tip}^2}$, with $\Delta_{tip} = \frac{1}{2} 3.5 K_B T$. Here, $\Delta$ is the HWHM obtained from the fit to a Fano function and $w_{eff}$ is the effective width.

In a second step we correct for the broadening due to the lock-in oscillation by defining an effective temperature as proposed by Girovsky et al. [3]: $T_{eff} = \frac{1}{5.4 K_B} \sqrt{(5.4 K_B T)^2 + (1.7 V_{rms})^2}$ . Here, $T_{eff}$ is the effective temperature, $T$ is the measurement temperature and $V_{rms}$ is the root mean square value of the lock-in oscillation voltage.

In a last step, the Kondo temperature $T_K$ is extracted fitting the effective width as a function of the effective temperature to the Fermi-Liquid model [4]: $w_{eff} = \frac{1}{2}\sqrt{(\alpha K_B T_{eff})^2 + (2K_B T_K)^2}$. The α parameter extracted from the fit is 7.4±0.3.